\documentstyle{article}

\def\be{\begin{equation}}
\def\ee{\end{equation}}
\def\bea{\begin{eqnarray}}
\def\eea{\end{eqnarray}}
\def\({\left(}
\def\){\right)}

\begin{document}

\pagestyle{empty}
\vskip-10pt
\vskip-10pt
\hfill {\tt hep-th/0301220}
\begin{center}
\vskip 3truecm
{\Large\bf
Low energy dynamics of self-dual $A_1$ strings
}\\ 
\vskip 2truecm
{\large\bf
Andreas Gustavsson
}\\
\vskip 1truecm
{\it Institute of Theoretical  Physics,
Chalmers University of Technology, \\
S-412 96 G\"{o}teborg, Sweden}\\
\vskip 5truemm
{\tt f93angu@fy.chalmers.se}
\end{center}
\vskip 2truecm
\noindent{\bf Abstract:} 
We examine the interrelation between the (2,0) supersymmetric six dimensional effective action for the $A_1$ theory, and the corresponding low-energy theory for the collective coordinates associated to selfdual BPS strings. We argue that this low energy theory is a two-dimensional N = 4 supersymmetric sigma model.
\vfill \vskip4pt

\eject
\newpage
\pagestyle{plain}
 
\section{Introduction}
We do not have a microscopic understanding of the interacting $(2,0)$ supersymmetric quantum theories in six dimensions, although many years now have passed since they where first discovered by Witten in 1995 \cite{Witten}. But whatever the microscopic theory is, we may always consider the low-energy effective action obtained by integrating out all massive degrees of freedom, provided that we have given a non-zero vacuum expectation value to say the fifth scalar field in the $(2,0)$-supermultiplet, so that the selfdual strings and all other subtle degrees of freedom become massive. We will in this letter only consider the $A_1$ effective theory in which we have just one $(2,0)$-supermultiplet, which on-shell consists of a selfdual two-form gauge potential $B^+$ with selfdual field strength $H^+ = dB^+$, five scalars $\phi^a$, and a symplectic Majorana spinor $\psi$ (we do not know the off-shell multiplet if any such exists).

The low energy dynamics of supersymmetric magnetic monopoles in superYang-Mills theories has been derived in \cite{Gauntlett} for $N=2$ SYM and in \cite{Blum} for $N=4$ SYM. A natural next step would be to examine the low energy dynamics of supersymmetric selfdual string solitons in the $(2,0)$ supersymmetric theories in six dimensions. So far we have not derived the fully supersymmetric effective action for the $A_1$ theory. But already the terms we know should be in that action can be used to derive a formula for the moduli space metric (which requires only the kinetic terms in the effective action) and to derive that the moduli space must be a hyper-Kahler manifold. 

The microscopic theory is of course far out of reach at present, contrary to the situation for superYang-Mills theories. We know that for instance the hedgehog monopole solutions in $SU(2)$ gauge theory can be gauge transformed so that only the massless scalar field is non-zero. The analogue in six dimensions to the fields that have acquired mass due to the non-zero vacuum expectation value of the Higgs scalar are not known. We think that they correspond to the selfdual strings, but we have no microscopic understanding of these objects. But in analogy with the four-dimensional case we will make the assumption that there is a 'gauge' in which the massive degrees of freedom (whatever they are) are zero and the string soliton solutions are entirely described in terms of a massless fields. We may use the $SO(5)$ R-symmetry to put $\phi^A = 0$ ($A=1,2,3,4$). Then for any $\frac{1}{2}$-BPS configuration we have the Bogomolnyi equation $H=*d\phi^5$ for the spatial components of $H$ and where $*$ is with respect to the transverse space to the string world-sheets. These results should hold in any interacting theory that possesses linear (2,0) supersymmetry as explained in \cite{Gustavsson}.

\section{Effective action and collective coordinates}
We will assume that we have flat six-dimensional Minkowski space and linear $(2,0)$ supersymmetry. To construct a Hamiltonian invariant under this supersymmetry we can anti-commute two supercharges given by
\be
Q = \frac{1}{6} \int d^5 x \gamma^{MNP}\gamma^0 H_{MNP} \psi + 2\int d^5 x \gamma^{M}\gamma^0 \sigma_a \partial_{M} \phi^a \psi.
\ee
Here $\gamma^M$ and $\sigma^a$ denote gamma matrices of $SO(1,5)$ and $SO(5)_R$ respectively and $H=dB+A$. If we take $A$ to be an external field we will get the free theory in which only the selfdual part of $H$ couples to $A$. To get the interacting $A_1$ theory we take $A = A(\phi)$ to be a function of the five scalars. This will bring in a new term in the anticommutator of two supercharges, arising from the commutator $[A(\phi),\dot{\phi}]$. We will content ourselves with construcing an action that is classically supersymmetric. Then commutators mean Poisson brackets and $[A,\dot{\phi}] \sim \frac{\partial A(\phi)}{\partial \phi}$. We take the gauge potential as in \cite{Gustavsson}\footnote{This gauge potential $A$ will get corrections if the five branes are not embedded as flat parallel planes in eleven dimensional space-time.}
\be
A(\phi)_{MNP} = \frac{\pi (\phi^5 - 2|\phi|)}{2V_4 |\phi|^3(\phi^5 - |\phi|)^2}\epsilon_{ABCD}\phi^A\partial_M \phi^B \partial_N \phi^C \partial_P \phi^D
\ee
Here $|\phi|:=(\phi^a \phi_a)^{\frac{1}{2}}$ and $A,B,... = 1,2,3,4$. $V_4 = \frac{8\pi^2}{3}$. Carrying out the above construction we get the following supersymmetric $A_1$ effective action (after a Legendre transformation of the Hamiltonian),
\bea
&&\frac{1}{2\pi}\int \(-\frac{1}{2}H \wedge *H - d\phi^a \wedge *d\phi^a + H \wedge A(\phi) \right.\cr
&&\left. + i\psi^T c\Omega \gamma^{M}\partial_M \psi + \frac{2\pi}{6V_4}i\psi^T c\Omega \epsilon_{abcde}\frac{1}{|\phi|^{5}}\phi^a\partial_M \phi^b\partial_N \phi^c\partial_P \phi^d \gamma^{MNP} \sigma^e \psi + ... \). \label{toy}
\eea
We have written $+...$ for terms involving four and more fermions, as well as terms that we have to add in order to achive invariance under $A \rightarrow A + d\Lambda$, $B\rightarrow B - \Lambda$. In particular we must add the term $-\tilde{A}$ where $d\tilde{A} = *dA - A \wedge dA$ \cite{Aharony}. This term should then also be supersymmetrized. But the action is still supersymmetric if we skip the terms in $+...$. This action is also invariant under the Lorentz group $SO(1,5)$ and under a global R-symmetry group $SO(5)_R$. To make the $SO(5)_R$ symmetry manifest we must make an integration by parts to get $B\wedge dA(\phi)$. The spinors transform in $(4,4)$ of $SO(1,5) \times SO(5)_R$ and are constrained by a symplectic Majorana condition
\be
\bar{\psi} = \psi^T c \Omega
\ee
where $\bar{\psi} := \psi^{\dag} \gamma^0$ and where $c$ and $\Omega$ are charge conjugation matrices of $SO(1,5)$ and $SO(5)$ respectively. We choose the conventions so that ${\gamma^{M}}^T = -c\gamma^M c^{-1}$ and ${\sigma_a}^T = \Omega \sigma_a \Omega^{-1}$. 

The moduli spaces we will consider are those of $k=1,2,...$ parallel BPS saturated selfdual strings. These moduli spaces should be given by ${\cal{M}}_k \times {\bf{R}}^4$ where the center of mass moduli space is ${\bf{R}}^4$ and corresponds to the four transverse directions to the (center of mass of the) strings. It seems very plausible that the dimension of the moduli space is $4 k$ (corresponding, in the classical picture of widely separated strings, to the four transverse coordinates of each string). But to prove this one has to count e.g the number of fermionic zero modes in a configuration with total charge $2\pi k$.
Parallel BPS-saturated strings are necessarily straight. To describe their motion tangent the moduli space we thus need just one parameter, which we will take to be the time-coordinate. But the generic string configuration is not a BPS configuration, and to describe such a configuration we also need a parameter running along the strings, which we take to be $x^5$. The tangent space to a BPS string configuration should thus be a two dimensional plane in the configuration space of all physically permitted configurations. This should be the natural generalization of \cite{Manton}. If we let $X^i$ ($i = 1,...,\dim {\cal{M}}_k$) denote the bosonic moduli parameters on ${\cal{M}}_k$ and assume that the BPS strings are aligned in the $x^5$ direction, then we make the following expansions,
\bea
\phi^a(x^0,...,x^5) & = & \phi^a(x^1,...,x^4,X^i(x^0,x^5))+...\cr
B_{MN}(x^0,...,x^5) & = & B_{MN}(x^1,...,x^4,X^i(x^0,x^5))+...\label{expansion}
\eea
for the bosonic fields. Here $\phi(x^1,...,x^4,X^i)$ are the static BPS solutions, which we assumed could be parametrized by $X^i$. The $+...$ involve terms that has to be added in order to satisfy the equations of motion to higher orders in $n = n_{\partial_t} + n_f$, where $n_{\partial_t}$ are the number of time derivatives and $n_f$ the number of fermions, following the logic of \cite{Harvey}. When we let the moduli parameters depend on $x^0, x^5$, they will be called collective coordinates. We will use the indices $\mu, \nu, ... = 0,5$ for the string world-sheets, and $I = 1,2,3,4$ for the transverse coordinates.

For a single string, $k=1$, the fermionic zero modes corresponds to broken supersymmetries $u_-$ which are constant spinors such that $\gamma^{05}\sigma_5 u_{\pm} = \pm u_{\pm}$. The zero modes are given by
\bea
\psi_0 = \gamma^{I} \partial_I \phi^5 u_-.
\eea 
That these really are zero modes is justified in the appendix. We could promote the moduli $u_-$ to fermionic collective coordinates $u_-$ just by letting them depend on $x^0,x^5$ and expand the fermion field as 
\be
\psi = \gamma^I \partial_I \phi^5(x^I,X^I(x^0,x^5)) u_-(x^0,x^5).
\ee
However we want to label the fermionic and the bosonic collective coordinates with the same label. That is, for $k = 1$ we want to convert $u_-$ into $\lambda^I$. $u_-$ transforms in the representation $(-\frac{1}{2},2,2') \oplus (+\frac{1}{2},2',2)$ of $SO(1,1) \times SO(4) \times SO(4)_R$. We may find one such $u_-$ such that $\lambda^I : = \sigma^5\gamma^I u_-$ constitute four linearly independent spinors. These will be four two-component Majorana spinors under $SO(1,1)$. But we may choose other representations for the $SO(4)$ gamma matrices. If we transform $\gamma^I \rightarrow \gamma^J J_{J}{}^I$ then the Clifford algerbra is preserved if and only if $J_{I}{}^{I'} J_{J}{}^{J'} \delta_{I'J'} = \delta_{IJ}$. Demanding $J$ to have real entries (which is needed in order to preserve the hermiticity property of the gamma matrices) we find that $J$ is a unit quaternion. We thus have the freedom to make the redefinition $\lambda^I \rightarrow \lambda^J J_J{}^I$ of the collective coordinates. 

For a general $k$ we make the following ansatz for the collective coordinates,
\be
\psi(x^M) = \partial_i \phi^5(x^I,X^i(x^0,x^5)) \sigma^5\lambda^i(x^0,x^5)
\ee
or with $\lambda^i$ replaced by ${\cal{J}}_j{}^i \lambda^j$ where ${\cal{J}}_i{}^j$ is the quarternion induced by $J_I{}^J$. 

Only for widely separated strings do we have the interpretation of $X^i$ as being four transverse coordinates of $k$ solitonic strings. We have no proof that $\psi$ given above for $\lambda$ a constant spinor is a zero mode other than for widely separated strings (with a separation much larger than $\phi^5(\infty)^{-\frac{1}{2}}$). But, as we will see in the next section, if we make this ansatz we get for instance a covariant derivative of $\lambda$ in the low-energy action just as one should expect. It seems very unlikely that there could be any other ansatz that also would give that result. The ansatz we make combines bosonic and fermionic zero modes in such a simple and natural way that it would be surprising if this formula did not give a fermionic zero mode for constant $\lambda$ in any point in the moduli space. So this is what we will assume, and we will get no contradionary result when we assume this, which strenghtens our belief that the ansatz for the fermionic collective coordinates is correct, not only for well separated strings but everywhere in the moduli space.

\section{Low energy dynamics}
We get the action for the collective coordinates by expanding the fields about the BPS configuration, inserting this expansion into the 6d Wilsonian low-energy effective action (\ref{toy}) and integrating over the transverse coordinates. 

When inserting any on-shell solution, $H$ becomes self-dual and hence $H\wedge *H = 0$. Inserting the static field configuration we get $A_{0MN}=0$. Then we use the equation of motion $dH = dA$ to get $\int H \wedge A = \int_{R^2 \times R_+} H \int_{S^3} A = \int_{R^2 \times R_+} *H \int_{S^3} H = \int d\phi^5 \wedge dx^0\wedge dx^5 \int_{S^3} H = \int dx^0 dx^5 \int d\phi^5 \wedge H = \int dx^0 dx^5 QD$ where $Q = 2\pi k$ and $D=\phi^5(\infty)$. Similarly we get, using the Bogomolnyi equation, that $\int_{R^4} d\phi^5 \wedge *_4 d\phi^5 = \int d\phi^5 \wedge H$. The total tension of the strings is $2QD$ with our conventions. Finally the two-fermion interaction term vanishes due to a factor $\gamma_I\int d^4 x x^I = 0$. Using these results when inserting the collective coordinate expansions into (\ref{toy}), we get the action
\be
\frac{1}{2\pi}\int dx^0 dx^5 \(G_{ij}\(\{X^k\}\) \(\eta^{\mu\nu}\partial_{\mu} X^i \partial_{\nu} X^j + i{\lambda^i}^T c\Omega \gamma^{\mu} D_{\mu} \lambda^j\) - 4\pi k D + ...\) \label{low}
\ee
where
\be
G_{ij} = \int d^4 x \partial_i \phi^5 \partial_j \phi^5\label{metric}
\ee 
is the metric on the moduli space and
\be
D_{\mu} \lambda^i = \partial_{\mu} \lambda^i + \Gamma^i_{kl}\partial_{\mu} X^k \lambda^l
\ee
where
\be
\Gamma^i_{jk} = G^{il}\int d^4 x (\partial_j \partial_k \phi^5) \partial_l \phi^5.
\ee
It is easily seen that $\Gamma^i_{jk}$ is the Christoffel symbol associated with the metric $G_{ij}$, so $D_{\mu}$ is a covariant derivative. The terms in $+...$ in (\ref{low}) arise from those in $+...$ in (\ref{toy}) and (\ref{expansion}). 

To get this metric we relied on the assumption that all the massive degrees of freedom can be put to zero in any string soliton solution. We think this is a plausible assumption given the analogy with the situation for the t'Hooft monopole in four dimensions where one can find a gauge in which only the massless scalar field is non-zero. The string solution we constructed in \cite{Gustavsson} also confirms that this is a valid assumption. A problem is that it is not sufficient that this assumption is valid only at far distances from the strings, since we integrated over the entire tranverse space to the strings to get the metric above. So what happens close to the strings can not be ignored. Furthermore the form of the metric would not cease to hold just because to strings happened to be close to each other since, as we already have said, the ansatz we made for the collective coordinates we make everywhere on the moduli space. Either the assumption we have made about the possibility of 'gauging' all the massive degrees of freedom to zero is correct everywhere, and then the metric is given by (\ref{metric}) everywhere on the moduli space. Or else it is wrong (incomplete) everywhere on the moduli space, and should be completed by adding some subtle degree of freedom that have become massive upon Higgsing. But as we get both a consistent low-energy theory (at least as far as we could go) as well as supersymmetry variations (to be shown in a moment) by assuming the form of the metric above, we feel quite confident that the metric on the moduli space is indeed given by (\ref{metric}) and that this formula is valid everywhere on the moduli space. 

The unbroken supersymmetry $\epsilon_+$ relates the bosonic and fermionic collective coordinates. The bosonic zero modes may be obtained by an unbroken supersymmetry variation as
\bea
\delta \phi^a & = & \epsilon^T_+ c\Omega \sigma^a \psi_0\cr
& = & \epsilon^T_+ c\Omega \sigma^a \sigma^5\lambda^i \partial_i \phi^5.
\eea
Expanding $\delta \phi^a = (\delta X^i) \partial_i \phi^a$, we see that in a BPS configuration where $\partial_i \phi^A = 0$ the right hand side must also vanish. Indeed it does identically (which can be seen by inserting $\gamma^{05}\sigma^5$ and first let it act on $\lambda^i_-$ and then on $\epsilon_+$). Taking $a=5$ we see that the bosonic collective coordiniates $X^i$ are related to the fermionic collective coordinates $\lambda^i$ as
\be
\delta_{\epsilon} X^i = \epsilon^T_+ c\Omega \lambda^i.
\ee
Similarly, by inserting the expansions
\bea
\phi & = & \phi(X=const) + \delta X^i \partial_i \phi(X=const) + Ordo((\delta X)^2)\cr
B & = & B(X=const) + \delta X^i \partial_i B(X=const) + Ordo((\delta X)^2)
\eea
into the (unbroken) supersymmetry variation
\be
\delta_{\epsilon} \psi  =  -i\(\frac{1}{24}\gamma^{MNP}H^+_{MNP} + \frac{1}{2}\gamma^{M}\sigma_a \partial_{M}\phi^a\)\epsilon_+
\ee
and using that in the BPS configuration (that is for the fields at $X=const$) this variation vanishes, we get only a contribution from term involving derivatives of $\delta X^i$. To first order this is
\be
\delta_{\epsilon} \psi = -i\gamma^{\mu}\sigma_a (\partial_{\mu} X^i) \partial_i \phi^a \epsilon_+.
\ee
Noting that $\partial_i \phi^A = 0$, we get
\be
\delta_{\epsilon}(\partial_i \phi^5 \lambda^i) = -i \gamma^{\mu} \partial_{\mu} X^i \partial_i \phi^5 \epsilon_+.
\ee
Multiplying by $\partial_j \phi^5$ and integrating over the transverse coordinates $x^i$ we get, using (\ref{metric}),
\be
\delta_{\epsilon} \lambda^i = -i\gamma^{\mu} \partial_{\mu}X^i \epsilon_+ - \Gamma^i_{jk} \delta_{\epsilon} X^k \lambda^j.
\ee
The 2d spinors are two-component Majorana spinors. These are the supersymmetries of a non-linear sigma model. The $SO(4)\times SO(4)_R$ spinor indices on $\epsilon$ and $\lambda^i$ are always trivially contracted and we can forget about them from the two-dimensional point of view.

We have four supersymmetries obtained by substituting ${\cal{J}}_j{}^i \lambda^j$ for $\lambda^i$. So the 2d action should be the $N=4$ supersymmetric sigma model. As shown in \cite{Alvarez-Gaume} the quaternions ${\cal{J}}_i{}^j$ comprise the three complex structures of a hyper-Kahler manifold. So we conclude that the moduli space ${\cal{M}}_k$ is a hyper-Kahler manifold.

We know that the $N = 4$ supersymmetric sigma model \cite{Alvarez-Gaume} contains an additional term which is proportional to
\bea
R_{ijkl}\bar{\lambda}^i \lambda^k \bar{\lambda}^j \lambda^l & = & 2\(\partial_{^[j}\partial_{_[k} G_{i^]l_]}
+ G_{mn}\Gamma^m_{k[i}\Gamma^n_{j]l}\) \bar{\lambda}^i \lambda^k \bar{\lambda}^j \lambda^l\cr
& = & 2\(\int \partial_k \partial_{[i} \phi^5 \partial_{j]} \partial_l \phi^5 
+ G^{pq} \int d^4 x \partial_p \phi^5 \partial_k \partial_{[i} \phi^5  \int d^4 y \partial_q \phi^5 \partial_{j]} \partial_l \phi^5 \) \bar{\lambda}^i \lambda^k \bar{\lambda}^j \lambda^l\cr
& = & 4 \int \partial_k \partial_{[i} \phi^5 \partial_{j]} \partial_l \phi^5\bar{\lambda}^i \lambda^k \bar{\lambda}^j \lambda^l+...\label{terms}
\eea
where in the last step we have used the completeness and orthonormality conditions of the modes to rewrite the second term. To get this term we would need the 6d effective action up to four-fermion interactions (which we do not have yet). This could presumably be done easier once a superfield formalism has been developed for the tensor multiplet.

\newpage
\appendix

\section{Fermionic zero modes}
The equations of motion for the fermions associated to a string solitons aligned in the 5-direction read ($I = 1,2,3,4$)
\be
\gamma^{I}\partial_{I} \psi = -\frac{2\pi}{6V_4}\frac{1}{r^5}\epsilon^{IJKL}\epsilon_{abcde}\phi^a \partial_I \phi^b \partial_J \phi^c \partial_K \phi^d \sigma^e \gamma^{05}\gamma_L \psi
\ee
where will let $\rho = \sqrt{\phi^A \phi_A}$ ($A=1,...,4$) and $r=\sqrt{\rho^2 + (\phi^5)^2}$. 

For $k = 1$ the map $x^A \mapsto \phi^A$ has winding number one, and $\phi^A = \rho(R) \frac{x^A}{R}$. We can then write the equation of motion as
\be
\gamma^{I}\partial_{I} \psi = \frac{2\pi}{V_4}\(\frac{\rho^4}{r^5}\frac{x^I}{R^4}\gamma_I\gamma^{05} \sigma^5 + ...\)\psi
\ee
Here $+...$ are terms proportional to $\gamma_A\gamma^{05}\sigma^A \psi$ and to $x_A\gamma^{05}\sigma^A \psi$.

Inserting that $\psi_{\pm} = \gamma^{I}\partial_I \phi^5 u_{\pm}$ where the BPS solution (\cite{Gustavsson}) is given by $\phi^5 = (D - \frac{|Q|}{2V_3 R^2})\theta(R - R_0)$ with $Q = 2\pi$ and $R_0 = \sqrt{|Q|/(2V_3 D)}$, we get the l.h.s.
\be
\gamma^{J} \partial_{J}(\gamma^I \partial_I \phi^5 u_{\pm}) = (\partial^I \partial_I \phi^5) u_{\pm} = \frac{|Q|}{V_3 {R_0}^3}\delta(R - R_0)u_{\pm}
\ee
To compute the r.h.s. we need the following results,
\bea
\lim_{\rho \rightarrow 0}\frac{\rho^4}{r^5} & = & \frac{4}{3}\delta(\phi^5)\cr
\delta(\phi^5) & = & \frac{\delta(R - R_0)}{\frac{d\phi^5}{dR}(R_0)} = \frac{V_3{R_0}^3}{|Q|} \delta(R-R_0)
\eea
The terms $+...$ should combine into something that is proportional to $\delta'(\phi^5)\phi^5 + \delta(\phi^5) = \(\delta(\phi^5)\phi^5\)'$  (with $'$ denoting $d/d\phi^5$) and thus vanishes identically. We then get the r.h.s.
\bea
- \frac{2\pi}{V_4}\frac{\rho^4}{r^5}\frac{x^I}{R} \gamma_I \gamma^{05} \sigma^5 \psi_{\pm} & = & \mp \frac{2\pi}{V_4}\frac{4}{3}\delta(\phi^5)\frac{x^I}{R^4}\gamma_I \psi_{\pm}\cr
& = & \mp \frac{|Q|}{V_3 {R_0}^3}\delta(R - R_0) u_{\pm}
\eea
The r.h.s. is then equal to the l.h.s. only for the choice $u_-$. These are thus fermionic moduli for $k=1$.

\end{document}